\documentstyle[aps,epsf]{revtex}

\begin{document}
%\twocolumn[\hsize\textwidth\columnwidth\hsize\csname @twocolumnfalse\endcsname
\title{Electronic structure of spheroidal fullerenes in a weak uniform magnetic field: a continuum field-theory model}
\author{M. Pudlak$^a$, R. Pincak$^{a,b}$ and V.A. Osipov$^b$}
\address{
$^a$Institute of Experimental Physics, Slovak Academy of Sciences,
Watsonova 47,043 53 Kosice, Slovak Republic\\
$^b$Joint Institute for Nuclear Research, Bogoliubov Laboratory of
Theoretical Physics,
141980 Dubna, Moscow region, Russia\\
e-mail: pudlak@saske.sk, pincak@saske.sk, osipov@thsun1.jinr.ru }
\address{\em (\today)}
\preprint \draft \maketitle
\begin{abstract}
The effect of a weak uniform magnetic field on the electronic
structure of slightly deformed fullerene molecules is studied within
the continuum field-theory model. It is shown how the existing due
to spheroidal deformation fine structure of the electronic energy
spectrum modifies in the presence of the magnetic field. Exact
analytical solutions for zero-energy modes are found.

\end{abstract}

\vskip 0.1cm \pacs{PACS numbers: 36.40.Cg, 33.55.Be, 71.20.Tx}
\vskip 0.1cm

%\vskip 0.1cm

%The energy shift of the electronic spectra of fullerenes with
%elliptical geometries within a gauge field-theory were predicted.

\section{Introduction}

Recently, we have considered the problem of the low energy
electronic states in spheroidal fullerenes~\cite{Pudlak}. The main
findings were a discovery of fine structure with a specific shift of
the electronic levels upwards due to spheroidal deformation. In
addition, three twofold degenerate modes near the Fermi level with
one of them being the true zero mode were found. An interesting
question is the modification of this structure under the influence
of a uniform magnetic field.

The problem of Zeeman splitting and Landau quantization of electrons on
a sphere was studied in Ref.~\cite{Aoki}. To this end, the Schr\"odinger
equation for a free electron on the surface of a sphere in a uniform magnetic
field was formulated and solved. In this paper, we explore the field-theory
model suggested in Ref.~\cite{Pudlak}, which describes the electronic states near
the Fermi energy and takes into account the specific structure of carbon lattice,
geometry, and the topological defects (pentagons). The Euler's theorem
for graphene requires the presence of twelve pentagons to get the
closed molecule. In the framework of continuum description we
extend the Dirac operator by introducing the Dirac monopole field
inside the spheroid to simulate the elastic vortices due to twelve
pentagonal defects. The K spin flux which describes the exchange
of two different Dirac spinors in the presence of a conical
singularity is included in a form of t'Hooft-Polyakov monopole.
Our studies cover slightly elliptically deformed molecules
in the weak uniform external magnetic field.

\section{The model}

Let us start with writing down the Dirac operator for free massless
fermions on the Riemannian spheroid $S^{2}$.
%To incorporate fermions on the curved background, we need a set of
%orthonormal frames $\{e_{\alpha}\}$, which yield the same metric,
%$g_{\mu\nu}$, related to each other by the local $SO(2)$ rotation,
%$$e_{\alpha}\to e'_{\alpha}={\Lambda}_{\alpha}^{\beta}e_{\beta},\quad
%{\Lambda}_{\alpha}^{\beta}\in SO(2).$$
%It then follows that
%$g_{\mu\nu} = e^{\alpha}_{\mu}e^{\beta}_{\nu} \delta_{\alpha \beta}$
%where $e_{\alpha}^{\mu}$ is the zweibein, with the orthonormal frame
%indices being $\alpha,\beta=\{1,2\}$, and the coordinate indices
%$\mu,\nu=\{1,2\}$. As usual, to ensure that physical observables are
%independent of a particular choice of the zweibein fields, a local
%$so(2)$ valued gauge field $\omega_{\mu}$ is to be introduced. The
%gauge field of the local Lorentz group is known as a spin
%connection.
The Dirac equation on a surface $\Sigma$ in the presence of the
abelian magnetic monopole field $W_{\mu}$ and the external
magnetic field $A_{\mu}$ is written as~\cite{Davies}
\begin{equation}
i\gamma^{\alpha}e_{\alpha}^{\ \mu}[\nabla_{\mu} - iW_{\mu}-
iA_{\mu}]\psi = E\psi, \label{eq:1}
\end{equation}
where $e_{\alpha}^{\mu}$ is the zweibein,
$g_{\mu\nu} = e^{\alpha}_{\mu}e^{\beta}_{\nu} \delta_{\alpha \beta}$ is the metric,
the orthonormal frame indices $\alpha,\beta=\{1,2\}$, the coordinate indices
$\mu,\nu=\{1,2\}$, and $\nabla_{\mu}=\partial_{\mu}+\Omega_{\mu}$ with
\begin{equation}
\Omega_{\mu}=\frac{1}{8}\omega^{\alpha\ \beta}_{\ \mu}
[\gamma_{\alpha},\gamma_{\beta}] \label{eq:2},
\end{equation}
being the spin connection term in the spinor representation (see
\cite{Pudlak} for details). The energy in (\ref{eq:1}) is measured
from the Fermi level.

%The Fermi level is situated in the middle of the well-know HOMO-LUMO
%gap where HOMO is highest occupied molecular orbital and LUMO is
%lowest unoccupied molecular orbital.

\subsection{Zero-energy mode}

Let us start from the analysis of an electron state
at the Fermi level (so-called zero-energy mode).
We will use the projection coordinates in the form
\begin{equation}
x=\frac{2R^{2}r}{R^{2}+r^{2}} \cos\phi; \quad
y=\frac{2R^{2}r}{R^{2}+r^{2}} \sin\phi; \quad
z=-c\frac{R^{2}-r^{2}}{R^{2}+r^{2}},\label{eq:3}
\end{equation}
where $R$ and $c$ are the spheroidal axles.
The Riemannian connection with respect to the orthonormal frame is
written as~\cite{Nakahara,Gockeler}
\begin{equation}
-\omega{_{\phi 2}^{1}}=\omega{_{\phi
1}^{2}}=\frac{R^{2}-r^{2}}{\sqrt{(R^{2}-r^{2})^{2}+4c^{2}r^{2}}}\
; \quad \omega{_{r 2}^{1}}=\omega{_{r 1}^{2}}=0. \label{eq:4}
\end{equation}

The only nonzero component of the gauge field $W_{\mu}$ in region
$R_{N}$  for spheroidal fullerenes reads (see Ref.~\cite{Pudlak})
\begin{equation}
W_{\phi}=G+m\frac{z}{X},\label{eq:5}
\end{equation}
where
\begin{equation}
X=\frac{\sqrt{c^{2}(r^2-R^2)^2+4R^{4}r^2}}{r^2+R^2}.
\label{eq:6}
\end{equation}
Notice that the monopole field $W_{\mu}$ in Eq. (\ref{eq:5})
consists of two parts. The first one comes from the K-spin
connection term and implies the charge $g=\pm 3/2$ while the
second one is due to elastic flow through a surface. This
contribution is topological in its origin and characterized by
charge $G$. For total elastic flux from twelve pentagonal defects
one has $G=1$. The parameter $m=g-G$ is introduced in Eq.(\ref{eq:5}).
The uniform external magnetic field $B$ is chosen
to be pointed in the $z$ direction so that $
\vec{A}=B\left(y,-x,0\right)/2$. In projection coordinates the
only nonzero component of $A_\mu$ is written as
\begin{equation}
A_{\phi}=-\frac{2BR^{4}r^{2}}{(R^{2}+r^{2})^{2}}.\label{eq:7}
\end{equation}
Let us study Eq.(\ref{eq:1}) for the electronic states at the Fermi energy ($E=0$).
The Dirac matrices can be chosen to be the Pauli matrices, $\gamma_1=-\sigma_2,
\gamma_2=-\sigma_1$. By using the substitution
\begin{equation}
\left(%
\begin{array}{c}
  \psi_{A} \\
  \psi_B \\
\end{array}%
\right) =\sum_j \frac{e^{i(j+G)\phi}}{\sqrt{2\pi}}\left(%
\begin{array}{c}
  u_j(r) \\
  v_j(r) \\
\end{array}%
\right) ,j=0,\pm 1,\pm 2,\ldots
\label{eq:8}
\end{equation}
we obtain
\begin{equation}
\left(\frac{1}{r}(j-m\frac{z}{X}-A_{\phi})+\frac{R^2+r^2}{K}(\partial_{r}-\frac{r^2-R^2}{2r(r^2+R^2)})\right)v(r)=0,
\label{eq:9}
\end{equation}
\begin{equation}
\left(\frac{1}{r}(j-m\frac{z}{X}-A_{\phi})-\frac{R^2+r^2}{K}(\partial_{r}-\frac{r^2-R^2}{2r(r^2+R^2)})\right)u(r)=0,
\label{eq:10}
\end{equation}
where $K=\sqrt{(R^2-r^2)^2+4c^{2}r^2}$.
We assume that the eccentricity of the spheroid is small enough. In
this case, one can write down $c=R+\delta R$ and $\delta$
($\mid\delta\mid\ll1$) is a small dimensionless parameter
characterizing the spheroidal deformation. Therefore, one
can follow the perturbation scheme using $\delta$ as the
perturbation parameter. To the leading in $\delta$ approximation
Eqs. (\ref{eq:11}) and (\ref{eq:12}) are written as
$$\partial_{r}v_{j}(r)=(-\frac{(R^2+r^2)^2}{r[(R^2+r^2)^2-4R^{2}r^{2}\delta]}[j-m(1+\delta)\frac{r^2-R^2}{r^2+R^2}+
m\delta\frac{(r^2-R^2)^{3}}{(r^2+R^2)^{3}}]$$
\begin{equation}
-\frac{2BR^{4}r}{[(R^2+r^2)^2-4R^{2}r^{2}\delta]}+\frac{r^2-R^2}{2r(R^2+r^2)})v_{j}(r),
\label{eq:11}
\end{equation}
$$\partial_{r}u_{j}(r)=(\frac{(R^2+r^2)^2}{r[(R^2+r^2)^2-4R^{2}r^{2}\delta]}[j-m(1+\delta)\frac{r^2-R^2}{r^2+R^2}
+m\delta\frac{(r^2-R^2)^{3}}{(r^2+R^2)^{3}}]$$
\begin{equation}
+\frac{2BR^{4}r}{[(R^2+r^2)^2-4R^{2}r^{2}\delta]}+\frac{r^2-R^2}{2r(R^2+r^2)})u_{j}(r).
\label{eq:12}
\end{equation}
The exact solution to Eqs. (\ref{eq:11}) and (\ref{eq:12}) is found to be
\begin{equation}
v_{j}(x)=\frac{(x+1)^{(1/2-m)}}{x^{\frac{1}{2}(j+m+1/2)}}
\left(\frac{x+1-2\delta-2\sqrt{\delta(\delta-1)}}
{x+1-2\delta+2\sqrt{\delta(\delta-1)}}\right)^{-\alpha}\left((x+1)^{2}-4\delta
x\right)^{m}, \label{eq:13}
\end{equation}
\begin{equation}
u_{j}(x)=(x+1)^{(m+1/2)} x^{\frac{1}{2}(j+m-1/2)}
\left(\frac{x+1-2\delta-2\sqrt{\delta(\delta-1)}}
{x+1-2\delta+2\sqrt{\delta(\delta-1)}}\right)^{\alpha}\left((x+1)^{2}-4\delta
x\right)^{-m}, \label{eq:14}
\end{equation}
where $x=r^2/R^2
$,
$\alpha=(j/2)\sqrt{\delta/(\delta-1)}+BR^{2}/4\sqrt{\delta(\delta-1)}
$.
The function $u_{j}$ can be normalized if the condition
$m-1/2<j<1/2-m$ is fulfilled. In turn, the normalization condition
for $v_{j}$ has the form $-1/2-m<j<m+1/2$. Finally, there are five
normalized solutions for $u_{j}$ with $j=0,\pm 1,\pm 2$ when
$m=-5/2$, and one solution for $v_{j}$ with $j=0$ when $m=1/2$. In
the limit $\delta\rightarrow 0$ we arrive at the solution for
spherical fullerenes
\begin{eqnarray}
v_j(r) = r^{(-1/2-j-m)}(R^{2}+r^{2})^{(1/2+m)}e^{\frac{BR^{4}}{R^{2}+r^{2}}},\nonumber\\
u_j(r) = r^{(-1/2+j+m)}(R^{2}+r^{2})^{(1/2-m)}e^{-\frac{BR^{4}}{R^{2}+r^{2}}}.\label{equv}
\label{eq:15}
\end{eqnarray}
Since normalization conditions do not depend on the parameter $\delta$
they are the same as for spherical case.
Thus the perturbation does not change the number of zero modes.
Notice that Eq. (\ref{eq:15}) can be obtained
from Eqs. (\ref{eq:13}) and (\ref{eq:14}) by using $\lim\left[1+(x/a)\right] ^{a}=e^{x}, a\rightarrow\infty$.

\subsection{Electronic states near the Fermi energy}

Let us study now the structure of electron levels near the Fermi energy.
It is convenient to consider this problem by using of the Cartesian
coordinates $x$, $y$, $z$ in the form
\begin{equation}
x=a~\sin\theta \cos\phi; \quad y=a~\sin\theta \sin\phi; \quad
z=c~\cos\theta. \label{eq:16}
\end{equation}
The Riemannian connection reads (cf. (\ref{eq:4}))
\begin{equation}
\omega{_{\phi 2}^{1}}=-\omega{_{\phi 1}^{2}}=\frac{a
\cos\theta}{\sqrt{a^{2}\cos^{2}\theta+c^{2}\sin^{2}\theta}}\ ; \quad
\omega{_{\theta 2}^{1}}=\omega{_{\theta 1}^{2}}=0. \label{eq:17}
\end{equation}
Within the framework of the perturbation scheme the spin connection coefficients
are written as
\begin{equation}
\omega{_{\phi 2}^{1}}=-\omega{_{\phi
1}^{2}}\approx\cos\theta(1-\delta\sin^{2}\theta), \label{eq:18}
\end{equation}
where the terms to first order in $\delta$ are taken into account.
In spheroidal coordinates, the only nonzero component of $W_{\mu}$
in region $R_{N}$ is found to be (see Ref.~\cite{Pudlak})
\begin{equation}
W_{\phi}\approx g\cos\theta(1+\delta\sin^{2}\theta) +
G(1-\cos\theta)-\delta G\sin^{2}\theta
\cos\theta,\label{eq:19}\end{equation}
and the external magnetic field reads
\begin{equation}
A_{\phi}=-\frac{1}{2}Ba^{2}\sin^{2}\theta.\label{eq:20}
\end{equation}
By using the substitution (\ref{eq:8}) we obtain the
Dirac equation for functions $u_j$ and $v_j$ in the form
\begin{equation}
\left(-i\sigma_{1}\frac{1}{a}(\partial_{\theta}+\frac{\cot\theta}{2})+\frac{\sigma_{2}}{a
\sin\theta}\left(j-m\cos\theta+\frac{1}{2}Ba^{2}\sin^{2}\theta\right)+\delta
\hat{\cal{D}}_{1}\right) \left(\begin{array}{c}
  u_j(\theta) \\
  v_j(\theta) \\
\end{array}
\right)=E\left(\begin{array}{c}
  u_j(\theta) \\
  v_j(\theta) \\
\end{array}
\right), \label{eq:21}
\end{equation}
where
\begin{equation}
\hat{\cal{D}}_{1}=-\frac{\gamma_{1}}{a}\sin\theta\left(j-2m\cos\theta\right)-\gamma_{1}\frac{B
a}{2}\sin^{3}\theta. \label{eq:22}
\end{equation}
A convenient way to study the eigenvalue problem is to square Eq.
(\ref{eq:21}). For this purpose, let us write the Dirac operator in
Eq. (\ref{eq:21}) as
$\hat{\cal{D}}=\hat{\cal{D}}_{0}+\delta\hat{\cal{D}}_{1}$. One can
easily obtain that (similar way as in Ref.~\cite{Osipov})
\begin{eqnarray}
\hat{\cal{D}}_{0}^2=-\frac{1}{a^{2}}\left(\partial^{2}_{\theta}+\frac{\cos\theta}{\sin\theta}\partial_{\theta}-\frac{1}{4}-\frac{1}{4
\sin^{2}\theta}\right)+\frac{\left(j-m\cos\theta\right)^{2}}{a^{2}\sin^{2}\theta}+Bj
+\frac{B}{2}\left(\sigma_{3}-2m\right)\cos\theta \nonumber\\
+\sigma_{3}\frac{m-j\cos\theta}{a^{2}\sin^{2}\theta}+\frac{B^{2}a^{2}\sin^{2}\theta}{4}.
\label{eq:23}
\end{eqnarray}
Notice that to first order in $\delta$ the square of the Dirac
operator is written as $
\hat{\cal{D}}^{2}=(\hat{\cal{D}}^{2}_{0}+\delta\hat\Gamma), $ where
$
\hat\Gamma=(\hat{\cal{D}}_{0}\hat{\cal{D}}_{1}+\hat{\cal{D}}_{1}\hat{\cal{D}}_{0}).$
In an explicit form
$$
a^{2}\hat\Gamma=2j^{2}+jx\left(\sigma_{3}-6m\right)
+4m(m-\sigma_{3})x^{2}+2m\sigma_{3}+3Ba^{2}(1-x^{2})x(\sigma_{3}/2-m)
$$
\begin{equation}
 +2Bja^{2}(1-x^{2})+B^{2}a^{4}(1-x^{2})^{2}/2
\label{eq:24}
 \end{equation}
where the appropriate substitution $x=\cos\theta$ is used.
%\begin{eqnarray}
%a^{2}\hat\Gamma=-2j^{2}-j\left(\sigma_{3}-6m\right)\cos\theta-4m(m-\sigma_{3})\cos^{2}\theta-2m\sigma_{3}-3Ba^{2}\sin^{2}\theta\cos\theta\left(\frac{\sigma_{3}}{2}-m\right)\nonumber\\
%-2jBa^{2}\sin^{2}\theta-\frac{B^{2}a^{4}\sin^{4}\theta}{2}. \label{eq:24}
%\end{eqnarray}
The equation $\hat{\cal{D}}^2\psi=E^2\psi$ takes the form
$$
[\partial_x(1-x^2)\partial_x-\frac{(j-mx)^2-j\sigma_3
x+\frac{1}{4}+\sigma_3 m}{1-x^2}-a^{2}B V(x) -\delta a^{2}\hat\Gamma]\left(%
\begin{array}{c}
  u_j(x) \\
  v_j(x) \\
\end{array}%
\right)\nonumber
$$
\begin{equation}
=-(\lambda^2-\frac{1}{4})\left(%
\begin{array}{c}
  u_j(x) \\
  v_j(x) \\
\end{array}%
\right), \label{eq:25}
\end{equation}
where $\lambda=aE, V(x)=j+(\sigma_{3}-2m)x/2$. Since we consider the
case of a weak magnetic field the terms with $B^{2}$ and $\delta B$
in Eq. (\ref{eq:25}) can be neglected. As a result, the energy
spectrum for spheroidal fullerenes is found to be

\begin{equation}
(\lambda_{jn}^{\delta})^2=(n+|j|+1/2)^2-m^2+Ba^{2}j
+Ba^{2}A_{jn}+\delta F(j,n,m) \label{eq:26}
\end{equation}
where $$ A_{jn}=-\frac{j(m^2-1/4)}{p(p+1)}.$$

The function of spheroidal deformations $F(j,n,m)$ is specified in
Ref.~\cite{Pudlak}. Finally, in the linear in $\delta$
approximation, the low energy electronic spectrum of spheroidal
fullerenes takes the form
\begin{equation}
E_{jn}^{\delta}=E^0_{jn}+E^{0B_{z}}_{jn}+E^{\delta }_{jn}
\label{eq:29}
\end{equation}
with
\begin{eqnarray}
E^0_{jn}=\pm\sqrt{(2\xi+n)(2\eta+n)},\quad
E^{0B_{z}}_{jn}=\frac{Ba^{2}(j+A_{jn})}{2E^0_{jn}}, \nonumber
E^{\delta }_{jn}=\frac{\delta F(j,n,m)}{2E^0_{jn}}, \label{eq:30}
\end{eqnarray}
where $\xi=\mu\ (\nu)$ and $\eta=\beta\ (\alpha)$ for $j>0\  (j<0)$,
respectively. Here we came back to the energy variable $E=\lambda/a$
(in units of $\hbar V_F/a$ where $V_F$ is the Fermi velocity).

As is seen from Eq. (\ref{eq:29}), for $B=0$ the spheroidal
deformation gives rise to an appearance of a additional structure.
As an example, Table 1 shows all five contributions (in compliance
with Eq. (\ref{eq:29})) to the first and second energy levels for
YO-C$_{240}$ (YO means a structure given in \cite{yoshida,Lu}). As
is seen, the energy levels become shifted due to spheroidal
deformation. The uniform magnetic field provides the well-known
Zeeman splitting. The difference between topological and Zeeman
splitting is clearly seen. In the second case, the splitted levels
are shifted in opposite directions while for topological splitting
the shift is always positive. As an illustration, Fig. 1
schematically shows the structure of the second level. In this case,
the initial (for $\delta=0$) degeneracy of $E^0_{jn}$ is equal to
six. The spheroidal deformation provokes an appearance of three
shifted double degenerate levels (fine structure). The magnetic
field is responsible to Zeeman splitting.

We admit that the obtained values of the splitting energies can
deviate from the estimations within some more precise microscopic
lattice models and density-functional methods (see,
e.g.\cite{Broglia}). In our paper, we focused mostly on the very
existence of physically interesting effects. For this reason, the
values in both Table and the schematic picture are presented with
accuracy at about one percent of $E^{0}_{jn}$.

\section{Conclusion}

In conclusion, we have considered the structure of low energy
electronic states of spheroidal fullerenes in the weak uniform
magnetic field provided the spheroidal deformation from the sphere
is small enough. For the states at the Fermi level, we found an
exact solution for the wave functions. It is shown that the external
magnetic field modifies the density of electronic states and does
not change the number of zero modes. For non-zero energy modes,
electronic states near the Fermi energy of spheroidal fullerene are
found to be splitted in the presence of a weak uniform magnetic
field. It should be mentioned that the zero-energy states in our
model corresponds to the HOMO (highest occupied molecular orbital)
states in the calculations based on the local-density approximation
in the density-functional theory (see, e.g.~\cite{Saito}). In
particular, the HOMO-LUMO energy gap is found to be about $1.1$ eV
for YO-C$_{240}$ fullerene within our model (here LUMO means the
lowest unoccupied molecular orbital).

\vskip 0.2cm \vskip 0.2cm The work was supported in part by VEGA
grant 2/7056/27. of the Slovak Academy of Sciences, by the Science
and Technology Assistance Agency under contract No. APVT-51-027904
and by the Russian Foundation for Basic Research under Grant No.
05-02-17721.

\newpage

\begin{table}[htb]
\begin{center}
\begin{tabular}[textwidth]{l c c c c} \bfseries\bfseries $YO-C_{240}$\quad \quad $Ba^{2}=0.1$ & $j$ &
\bfseries $E^{0}_{jn} (eV)$ & \bfseries $ E^{0B_z}_{jn} (meV)$ &
 \bfseries $E^{\delta }_{jn}(meV)$\\
\hline\hline\bfseries $n=0$, $m=1/2$
&1&1.094&27&10.5\\
&-1&1.094&-27&10.5\\
\hline\hline \bfseries $n=1$, $m=1/2$
&1&1.89&16&8.8\\
&-1&1.89&-16&8.8\\
\bfseries  $n=0$, $m=1/2$
&2&1.89&32&28.4\\
&-2&1.89&-32&28.4\\
\bfseries $n=0$, $m=-5/2$
&3&1.89&24&3\\
&-3&1.89&-24&3\\
\end{tabular}
\end{center}
\caption{{\footnotesize The structure of the first and higher energy
levels for YO-C$_{240}$ fullerene in uniform magnetic field. The
hopping integral and other parameters are taken to be $t=2.5\ eV$
and $V_F=3t\overline{b}/2\hbar$, $\overline{b}=1.45{\textmd{\AA}}$,
$\overline{R}=7.03{\textmd{\AA}}$, $SD=0.17{\textmd{\AA}}$,
$\delta=0.024$. $\overline{b}$ is average bond length,
$\overline{R}\ (\overline{R}=a)$ is average radius, $SD$ is standard
deviation from a perfect sphere (see Refs. [7,8]), so that
$\delta=SD/\overline{R}$.}} \label{tab}
\end{table}

\begin{figure}[!ht]
\begin{center}
\epsfysize=6cm \epsffile{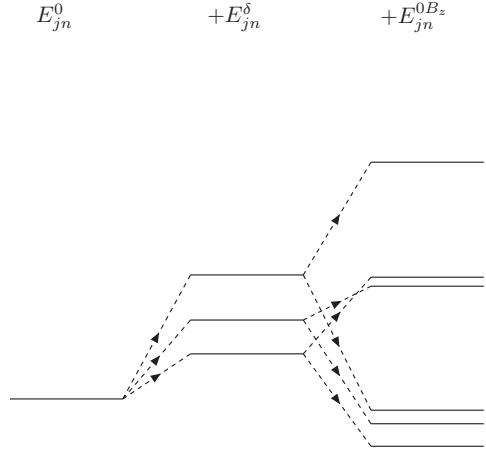}\epsfysize=4cm \caption[]{The
schematic picture of the second positive electronic level
$E_{jn}^{\delta}$ of the spheroidal fullerenes in a weak uniform
magnetic field.}
%\label{DOS123}
\end{center}
\end{figure}

\end{document}